# Two-micron wavelength high speed photodiode with InGaAs/GaAsSb type-II multiple quantum wells absorber

YAOJIANG CHEN,[1,2,3] ZHIYANG XIE,[1] JIAN HUANG,[1] ZHUO DENG,[1] BAILE CHEN,[1,*]

[1] *School of Information Science and Technology, ShanghaiTech University, Shanghai 201210, China*
[2] *Shanghai Institute of Microsystem and Information Technology, Chinese Academy of Sciences, Shanghai 200050, China*
[3] *University of Chinese Academy of Sciences, Beijing 100049, China*
*\*Corresponding author: chenbl@shanghaitech.edu.cn*

**Current optical communication system operating at 1.55 μm wavelength band may not be able to continually satisfy the growing demand on the data capacity within the next few years. Opening a new spectral window at around 2 μm wavelength with recently developed hollow-core photonic band gap fiber and thulium-doped fiber amplifier is a promising solution to increase the transmission capacity due to the low loss and wide bandwidth properties of these components at this wavelength. However, as a key component, the already demonstrated high speed photodetectors at 2 μm wavelength are still not comparable with those at 1.55 μm wavelength band, which chokes the feasibility of the new spectral window. In this work, we, for the first time, demonstrated a high speed uni-traveling carrier photodiode for 2 μm applications with InGaAs/GaAsSb type-II multiple quantum wells as the absorption region, which is lattice matched to InP. The device shows a 3dB bandwidth of 25 GHz at -3 V bias voltage and is, to the best of our knowledge, the fastest photodiodes among all group III-V and group IV photodetectors working in 2 μm wavelength range.**

## 1. Introduction

Due to the exponentially increasing volume of internet traffic, today's optical communication system are rapidly approaching their capacity limit [1,2]. This "capacity crunch" provides an impetus for developing next-generation optical communication system. A new spectral window at 2 μm wavelength is a promising solution to increase the system capacity due to the availability of low loss (0.2 dB/km) hollow-core photonic band gap fiber (HC-PBGF) and the high bandwidth (from 1810 to 2050 nm) thulium-doped fiber amplifier (TDFA) at 2 μm wavelength [3,4]. Studies on 2 μm optical components and system architecture have shown the promising potential of this new spectral window [5–7]. An eight-channel wavelength division multiplexing transmission system with 100 Gbit/s data capacity has been demonstrated at 2 μm band [6].

High speed photodetector is one of the key components of the optical communication system, and among various figure of mertis, bandwidth is most commonly used for high speed devices benchmark. At 2 μm wavelength band, high speed photodiodes have been demonstrated with In-rich InGaAs on InP [5,7–11], InGaAsSb on GaSb [12], GeSn/Ge on Si [13,14], and defect-mediated Si [15]. $In_{0.53}Ga_{0.47}As$ on InP, which cuts off at 1.7 μm, is the most widely used absorption material for high speed application due to its high carrier mobility. For 2 μm operation, In-rich InGaAs has to be used to extend the cut-off wavelength. Ye, et al. demonstrated a 10 GHz bandwidth photodiode with $In_{0.7}Ga_{0.3}As$ as absorption layer [11]. Joshi, et al. achieved a 16 GHz bandwidth using $In_{0.72}Ga_{0.28}As$ absorber [9]. Partially depleted photodiode with $Ga_{0.8}In_{0.2}As_{0.16}Sb_{0.84}$ absorber on GaSb has achieved a 3dB bandwidth of 6 GHz [12]. On the other hand, GeSn/Ge quantum well grown on Si shows benefit for intergration with other silicon devices, and a $Ge_{0.92}Sn_{0.08}$/Ge photodiode with 10 GHz bandwidth has been reported [14]. By inserting lattice defects, silicon on insulator photodiode can also operate at 2 μm wavelength and a 15 GHz bandwidth can be achieved [15]. However, the performance of these devices are sensitive to the crystal quality, making them inconvenient for practical application.

Short wavelength infrared (SWIR) photodetectors using InGaAs/GaAsSb type-II multiple quantum wells (MQWs) on InP substrate as absorber has been demonstrated and well studied recently [16–19]. The devices enjoy the advantage of lattice-matched property on InP, and show low dark current and high detectivity at room temperature. The crystal quality of this MQWs structure can be ensured since it is lattice-matched to InP and the growth technique of InP material system is mature. PIN based high speed photodiode with InGaAs/GaAsSb MQWs for 2 μm operation has achieved a 3dB bandwidth in the range of 3.5GHz to 10 GHz, which is mainly limited by the slow transport of optical-generated hole in the absorption region. [20–22].

To further improve the bandwidth, uni-traveling carrier photodiode (UTC-PD) with type-II MQWs absorber was proposed and theoretically investigated in our previous work [23]. UTC-PD based on InGaAs/InP material system has been demonstrated with hundreds of gigahertz bandwidth at 1.55 μm wavelength band for high speed application [24–27]. In UTC-PD structure, the optical-generated carriers are excited in the undepleted p-type absorption layer, and only electrons are injected into the InP drift layers, thus the effective carrier transit time is shorter than that of PIN structure with proper design.

In this work, we demonstrated normal incident InGaAs/GaAsSb MQWs uni-traveling carrier photodiodes at 2 μm wavelength band with a 3dB bandwidth of 25GHz and a bit rate of 30 Gbit/s. To the best of our knowledge, this is the highest bandwidth and bit rate demonstrated at 2 μm wavelength band. By analyzing the RF performance of the devices with different diameters, it is found that the 3dB bandwidth performance is still limited by RC-time constant, which can be further improved by optimizing the fabrication process.

## 2. Device structure and fabrication

The epi-layer structure of the photodiode is shown in Fig. 1(a). All layers are lattice-matched to InP substrate. The structure was grown on semi-insulating double-side-polished InP substrate by molecular beam epitaxy (MBE) system. The epitaxial growth began with 200 nm n+ InP layer, 20 nm n+ InGaAs layer and 900 nm n+ InP layer. The 100 nm n-doped InP layer with a doping concentration of $1\times10^{18}$ cm$^{-3}$ was then grown to reduce Si diffusion into the 400 nm intrinsic InP drift layer. The absorption layer consists of 180 nm graded doped 3nm/3nm InGaAs/GaAsSb MQWs. Following the MQWs layers are the 30 nm p-doped large band gap AlGaAsSb electron blocking layer to prevent the diffusing of electrons in absorption layer towards the p-doped InGaAs contact layer on the top. Room temperature photoluminescence (PL) of the epitaxial layers was conducted by using the 532nm wavelength laser as the excitation source and a Fourier transform infrared (FTIR) spectrometer with In-rich InGaAs detector. Two peaks can be identified in Fig. 1(b), indicating the good crystal quality of the material. The peak at 2.1 μm corresponds to the transition between the ground states in InGaAs and GaAsSb quantum well layers respectively, and the second peak at 1.75 μm may corresponds to the emission from the excited state in InGaAs or GaAsSb layer.

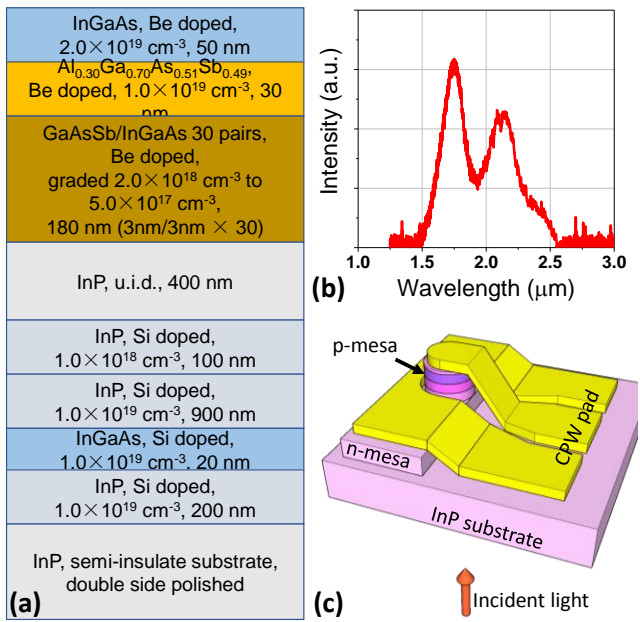

Fig. 1. (a) Epitaxial structure of the type-II MQWs UTC photodiode. (b) Photoluminescence measurement result of the epitaxial structure. (c) Schematic diagram of the fabricated device.

After the material growth, the epitaxial structure was processed into double mesa structures as shown in Fig. 1(c). A set of mesa devices with different diameters were formed by standard photolithography and wet etching process. A 150 μm pitch GSG coplanar waveguide (CPW) pad with an air bridge structure was electroplated for high frequency measurement. The substrate is backside polished to support backside illumination with no anti-reflction coating.

## 3. Measurement result

### A. Electrical characteristics

The dark current voltage (I-V) curves of the photodiodes at room temperature are shown in Fig. 2. The dark current value at -3 V is 3.02 nA, 3.88 nA, and 5.82 nA for the photodiodes with diameters of 10 μm, 20 μm, and 40 μm, respectively, which are lower than the In-rich InGaAs photodiodes [5,7–11]. Figure 3 shows the capacitance-voltage (C-V) curves. The device is fully depleted at -1 V, and the capacitance at -3 V is 60.6 fF, 154.0 fF, and 432.6 fF for the 10 μm, 20 μm, and 40 μm diameter photodiodes, respectively. Parasitic capacitance of 49.6 fF was found based on the linear fitting of capacitance v.s device area as shown in Fig. 4, which indicates that optimization is needed for future fabrication process.

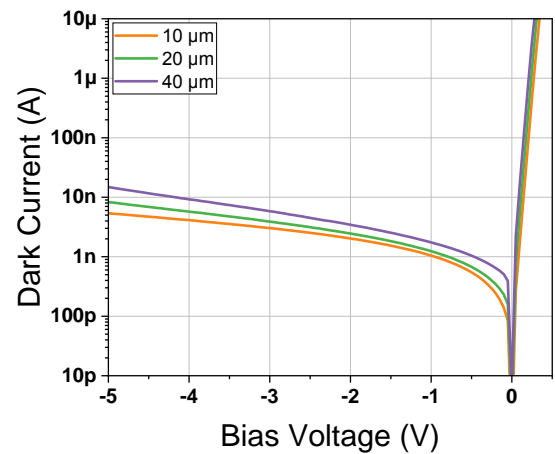

Fig. 2. Dark current characteristics for three devices with different diameters at room temperature.

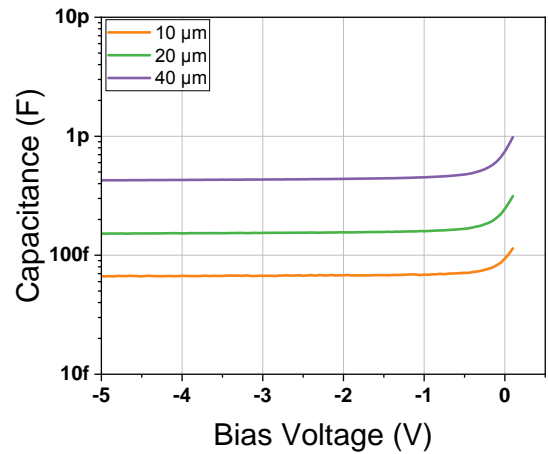

Fig. 3. Measured capacitance versus reverse bias for three devices with different diameters at room temperature.

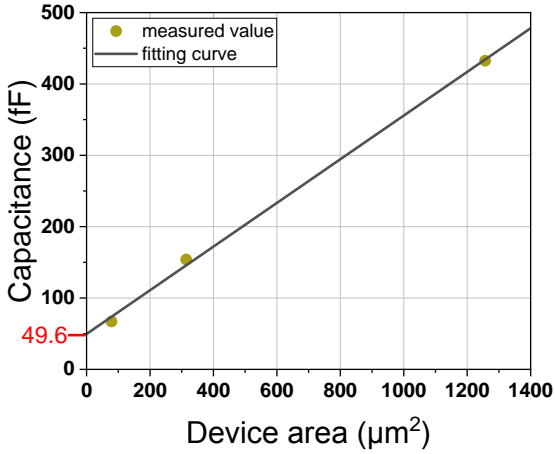

Fig. 4. Capacitance of devices at -3 V bias. The fitting result indicates a parasitic capacitance of 49.6 fF.

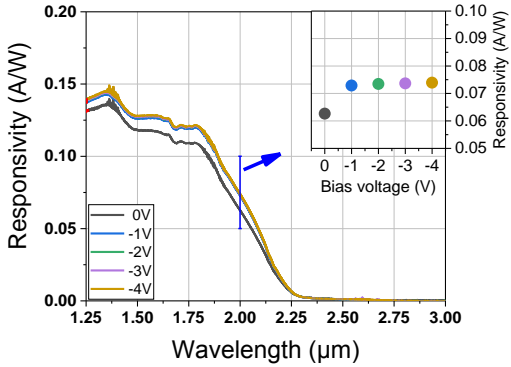

Fig. 5. Responsivity spectrum at various bias voltages. The inset shows the responsivity at 2 μm wavelength.

### B. Responsivity

The top-illuminated responsivity spectrum was measured using an FTIR spectrometer. And a standard blackbody radiation source at 700 °C was used to calibrate the responsivity. As shown in Fig. 5, the cut-off wavelength is larger than 2.25 μm, and the responsivity is 0.07 A/W at 2 μm wavelength and reverse bias (larger than -1 V). Considering the 180 nm thickness of absorption layer, the equivalent absorption coefficient of the absorption layer is about 3600 cm$^{-1}$.

### C. Bandwidth

The 3dB bandwidth was measured by a heterodyne setup with two beams of laser light at wavelength of 2004 nm coupled together, which generates an intensity-modulated light with frequency from hundreds of megahertz to more than 30 GHz. The modulated light was amplified and then illuminated the devices from backside.

Figure 6 shows the frequency response of a 10 μm diameter photodiode at different bias voltages. The 3dB bandwidth increases slightly while reverse bias increases. The highest 3-dB bandwidth of 25 GHz can be achieved at -4 V bias. Theoretically, the 3dB bandwidth of photodiode is often limited by transit time and RC time, as expressed by the equation:

$$f_{3dB} = \left(\sqrt{\frac{1}{f_T^2} + \frac{1}{f_{RC}^2}}\right)^{-1}, \qquad (1)$$

where $f_{3dB}$ is the total 3-dB bandwidth, $f_T$ is the transit time limit bandwidth, and $f_{RC}$ is the RC limit bandwidth. Recall the C-V measurement results in Fig. 4, the junction capacitance is almost constant once fully depleted (larger than -1 V reverse bias). As a result the small increase in 3-dB bandwidth should be caused by the enhancement in transit process when stronger electrical field is applied.

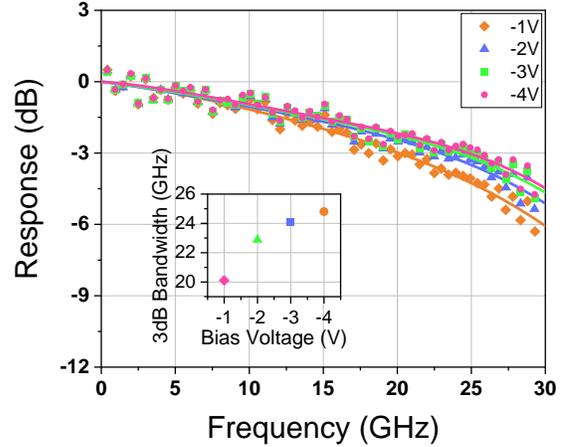

Fig. 6. Frequency response of a 10 μm photodiode at various bias voltages. The photocurrent is set to 1 mA. The inset shows the 3dB bandwidth at different bias voltages.

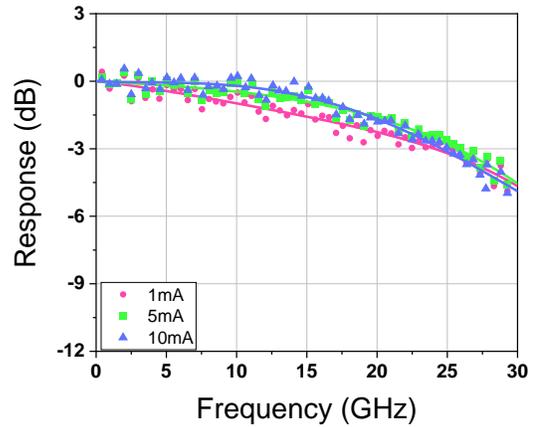

Fig. 7. Frequency response of a 10 μm photodiode at different photocurrents. The bias voltage is -3 V.

Figure 7 shows the frequency response of the 10 μm photodiode at different photocurrents. The 3-dB bandwidth is almost the same (25 GHz) when photocurrent increase from 1 mA to 10 mA, and no obvious saturation can be found.

Figure 8 shows the frequency response of three photodiodes with different diameters. The 3dB bandwidth of 40 μm, 20 μm, and 10 μm device is 5 GHz, 15 GHz, and 25 GHz, respectively. The smaller photodiode has smaller junction capacitance, leading to a larger RC limit bandwidth, while the transit limit bandwidth remains the same. Thus the strong bandwidth dependency on diameter indicates that the 3dB bandwidth of the devices are dominated by RC limit. To verify the RC limit, we measured the S-parameters and fitted the parameters with an equivalent circuit model as shown in Fig. 9(a). The fitting curves are shown in Fig. 9(b), 9(c), and 9(d). The extracted model parameters are listed in Table I. Then the theoretical RC limit frequency response can be calculated using the equivalent model, as shown in Fig. 9(e). The RC limit bandwidth of 40 μm, 20 μm, and 10 μm device is 7.7 GHz, 18.9 GHz, and 41.4 GHz, respectively. Applying equation (1), a rough estimation of the transit time limit bandwidth for the 10 μm diameter photodiode is about 31 GHz. As shown in Fig. 4 a parasitic capacitance of 49.6 fF exists in our devices, so the 3dB bandwidth of the devices can be further improved by reducing the parasitic capacitance. Figure 10 reviews the 3dB bandwidth of high speed photodetectors operating at 2 μm wavelength in recent years. With proper design, the InGaAs/GaAsSb MQWs UTC-PD can achieve excellent bandwidth performance than other devices.

regarded as open circuit. The measured and fitting curve of S11 of (b) 10 μm, (c) 20 μm, and (d) 40 μm photodiodes at -3 V bias (the blue curve is the measured data while the red curve is the fitting curve). (e) Calculated RC limit frequency response using the fitting results.

**Table I Fitting parameters**

| Device diameter | $R_s$ (Ohm) | $C_j$ (fF) | $L_s$ (pH) |
|---|---|---|---|
| 10μm | 13.4 | 67.8 | 30.1 |
| 20μm | 12.2 | 149.2 | 53.6 |
| 40μm | 4.7 | 404.9 | 77.1 |

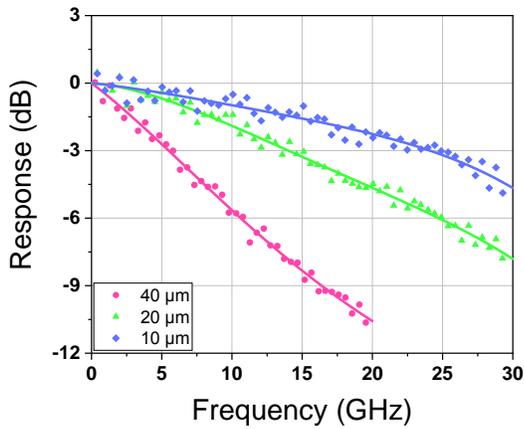

Fig. 8. Frequency response of photodiodes with different diameters. The bias voltage is -3 V and the photocurrent is 1 mA.

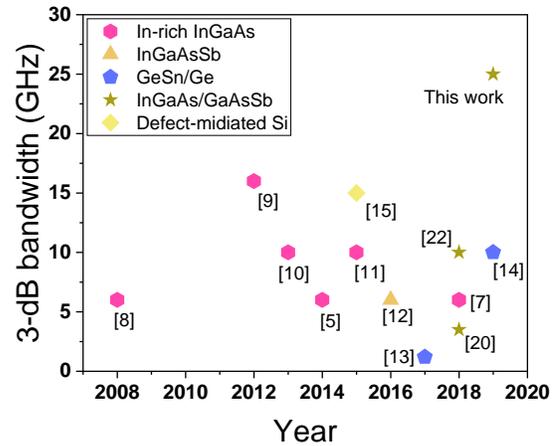

Fig. 10. A review of 3-dB bandwidth of high speed photodiodes operating at 2 μm wavelength reported in recent years.

Figure 11 shows the eye diagram of a 10 μm photodiode biased at -3 V and operating at 20 Gbit/s, 25 Gbit/s and 30 Gbit/s data rate. The 128 bits 2^15-1 PRBS sequences were generated as the data source to drive a 2 μm wavelength Mach-Zehnder modulator, which modulates the optical signal coming from a 2004nm wavelength single frequency fiber laser. The output of the photodiode was amplified by a +23 dB microwave amplifier and then displayed on the real-time sampling oscilloscope. Clear eye pattern is demonstrated at a 30 Gbit/s data rate, which indicates the devices can be used at 2 μm optical communications system.

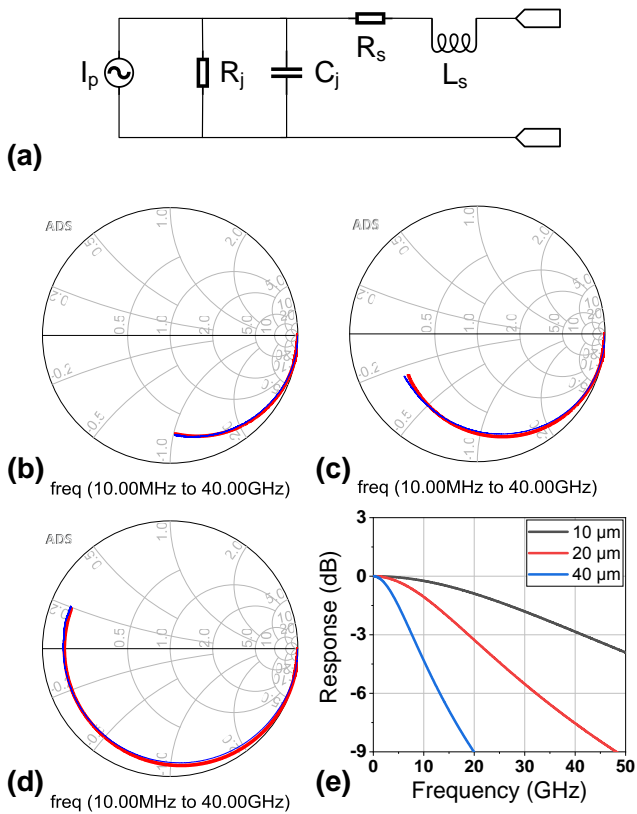

Fig. 9. (a) Equivalent circuit model used in parameter fitting. $C_j$ is the junction capacitance, $R_s$ is the series resistance (resistance of the ohm contacts and the CPW pads), $L_s$ is the inductance of the CPW pads, and $R_j$ is the junction resistance which is hundreds of mega-ohms and can be

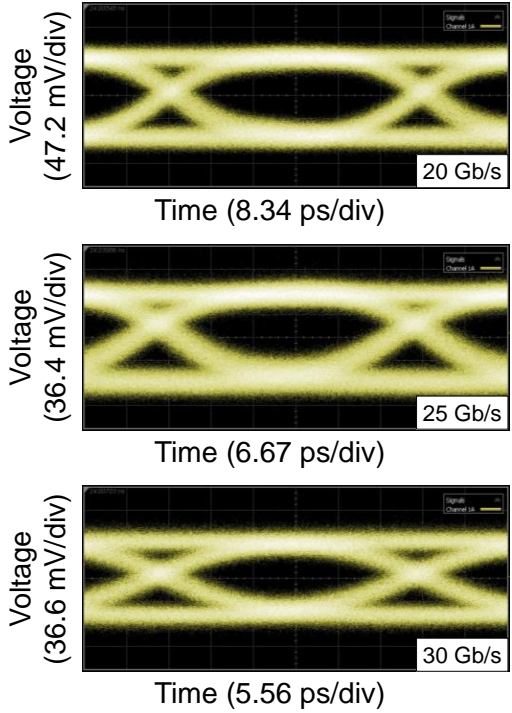

Fig. 11. Eye pattern of a 10 µm photodiode at 10 Gbit/s, 20 Gbit/s, 25 Gbit/s and 30 Gbit/s.

### D. Saturate power

For some applications such as microwave photonics, the high speed photodetector works in high power mode, and the saturate power is another important figure of merit. When the photocurrent is high, a large number of electrons exist in the depletion region, and the electrical field in the depletion region might collapse due to the space charge effect. This limits the carrier sweeping out process, causing a saturation in output power. Figure 12 and 13 shows the relationship between the output power and the photocurrent. In this measurement, lasers of 1.55 µm wavelength were used as optical source due to the power limitation of the 2 µm optical amplifier. The ideal output power of a photodetector can be calculated by:

$$P_{ideal} = m\frac{1}{2}I_p^2 R_L, \qquad (2)$$

where $I_p$ is the photocurrent, and $R_L$ is the load resistance, $m$ is the modulation depth which is almost 1 in our measurement. The compression in the figures is the difference between actual output power and the ideal output power. The 1dB compression point is defined as the point that the compression drops 1 dB compared to its maximal value, it describes the power handling ability of the devices. Table II listed the 1dB compression point of 10 µm and 20 µm devices at different bias voltage. The saturate power increases with the increase of the bias voltage, since the stronger electrical field can endure more space charge effect.

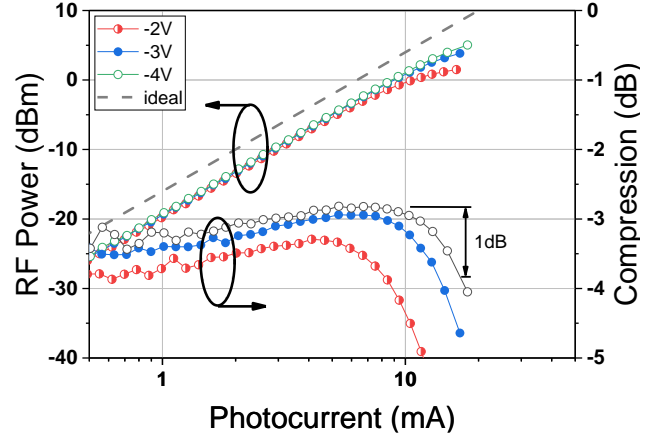

Fig. 12. Output RF power and compression versus photocurrent for a 10 µm photodiode at 25 GHz and at different bias voltages. The gray dash line shows the ideal output power.

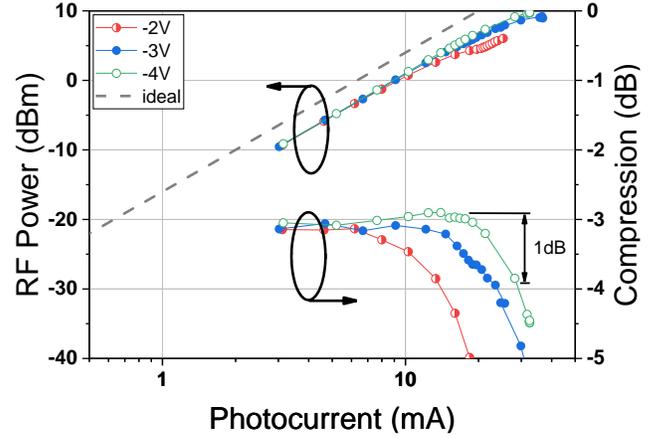

Fig. 13. Output RF power and compression versus photocurrent for a 20 µm photodiode at 15 GHz and at different bias voltages. The gray dash line shows the ideal output power.

**Table II 1dB compression points**

|       | -2V                  | -3V                  | -4V                  |
|-------|----------------------|----------------------|----------------------|
| 10 µm | -0.7 dBm (9.5 mA)    | 2.8 dBm (14.0 mA)    | 4.4 dBm (16.7 mA)    |
| 20 µm | 2.6 dBm (13.3 mA)    | 7.4 dBm (23.5 mA)    | 9.1 dBm (28.2 mA)    |

### 4. Conclusion

In this work, we have reported a high speed photodiode working at 2 µm wavelength based on InGaAs/GaAsSb type-II MQWs with uni-travelling carrier design. The InGaAs/GaAsSb type-II MQWs absorber cuts off at 2.25 µm, and the responisvity is 0.07 A/W at 2 µm. The 3-dB bandwidth at 2 µm is 25 GHz, 15 GHz, and 5 GHz for photodiode with 10 µm, 20 µm, and 40 µm diameter, respectively. A clear eye pattern can be observed at 30 Gb/s for the 10 µm photodiode. Analysis shows that the bandwidth of the current device is limited by RC limit. Therefore, future improvement of the device should be focused on optimization of the fabrication procedure to further reduce the parasitic capacitance.

**Funding**. Shanghai Sailing Program (17YF1429300); ShanghaiTech University startup funding (F-0203-16-002).

## REFERENCES

1. A. D. Ellis, J. Zhao, and D. Cotter, "Approaching the Non-Linear Shannon Limit," J. Light. Technol. **28**, 423–433 (2010).
2. D. J. Richardson, "Filling the Light Pipe," Science (80-. ). **330**, 327 LP-328 (2010).
3. P. J. Roberts, F. Couny, H. Sabert, B. J. Mangan, D. P. Williams, L. Farr, M. W. Mason, A. Tomlinson, T. A. Birks, J. C. Knight, P. St. J. Russell, and P. S. J. Russell, "Ultimate low loss of hollow-core photonic crystal fibres," Opt. Express **13**, 236 (2005).
4. Z. Li, A. M. Heidt, N. Simakov, Y. Jung, J. M. O. Daniel, S. U. Alam, and D. J. Richardson, "Diode-pumped wideband thulium-doped fiber amplifiers for optical communications in the 1800 -- 2050 nm window," Opt. Express **21**, 26450–26455 (2013).
5. B. Corbett, M. R. Gleeson, N. Ye, C. Robert, H. Yang, H. Zhang, N. Mac Suibhne, and F. C. G. Gunning, "InP-based active and passive components for communication systems at 2 μm," **33**, 4–6 (2014).
6. H. Zhang, N. Kavanagh, Z. Li, J. Zhao, N. Ye, Y. Chen, N. V. Wheeler, J. P. Wooler, J. R. Hayes, S. R. Sandoghchi, F. Poletti, M. N. Petrovich, S. U. Alam, R. Phelan, J. O'Carroll, B. Kelly, L. Grüner-Nielsen, D. J. Richardson, B. Corbett, and F. C. Garcia Gunning, "100 Gbit/s WDM transmission at 2 μm: transmission studies in both low-loss hollow core photonic bandgap fiber and solid core fiber," Opt. Express **23**, 4946 (2015).
7. F. G. Gunning, N. Kavanagh, E. Russell, R. Sheehan, and J. O. Callagha, "Key Enabling Technologies for Optical Communications at 2000 nm," (2018).
8. A. Joshi and D. Becker, "High-speed low-noise p-i-n InGaAs photoreceiver at 2-μm wavelength," IEEE Photonics Technol. Lett. **20**, 551–553 (2008).
9. A. Joshi and S. Datta, "High-speed, large-area, p-i-n InGaAs photodiode linear array at 2-micron wavelength," in *Infrared Technology and Applications XXXVIII* (2012), Vol. 8353, p. 83533D.
10. H. Yang, B. Kelly, W. Han, F. Gunning, B. Corbett, R. Phelan, J. O'Carroll, H. Yang, F. H. Peters, X. Wang, N. Nudds, P. O'Brien, N. Ye, and N. MacSuibhne, "Butterfly packaged high-speed and low leakage InGaAs quantum well photodiode for 2000nm wavelength systems," Electron. Lett. **49**, 281–282 (2013).
11. N. Ye, H. Yang, M. Gleeson, N. Pavarelli, H. Y. Zhang, J. O'Callaghan, W. Han, N. Nudds, S. Collins, A. Gocalinska, E. Pelucchi, P. O'Brien, F. C. G. Gunning, F. H. Peters, B. Corbett, J. O. Callaghan, W. Han, N. Nudds, S. Collins, E. Pelucchi, P. O. Brien, F. C. G. Gunning, F. H. Peters, and B. Corbett, "InGaAs Surface Normal Photodiode for 2 μm Optical Communication Systems," IEEE Photonics Technol. Lett. **4**, 1469–1472 (2015).
12. J. M. Wun, Y. W. Wang, Y. H. Chen, J. E. Bowers, and J. W. Shi, "GaSb-Based p-i-n Photodiodes with Partially Depleted Absorbers for High-Speed and High-Power Performance at 2.5-μm Wavelength," IEEE Trans. Electron Devices **63**, 2796–2801 (2016).
13. Y. Dong, W. Wang, S. Xu, D. Lei, X. Gong, X. Guo, H. Wang, S.-Y. Lee, W.-K. Loke, S.-F. Yoon, and Y.-C. Yeo, "Two-micron-wavelength germanium-tin photodiodes with low dark current and gigahertz bandwidth," Opt. Express **25**, 15818–15827 (2017).
14. S. Xu, W. Wang, Y.-C. Huang, Y. Dong, S. Masudy-Panah, H. Wang, X. Gong, and Y.-C. Yeo, "High-speed photo detection at two-micron-wavelength: technology enablement by GeSn/Ge multiple-quantum-well photodiode on 300 mm Si substrate," Opt. Express **27**, 5798 (2019).
15. J. J. Ackert, D. J. Thomson, L. Shen, A. C. Peacock, P. E. Jessop, G. T. Reed, G. Z. Mashanovich, and A. P. Knights, "High-speed detection at two micrometres with monolithic silicon photodiodes," Nat. Photonics **9**, 393–396 (2015).
16. B. Chen, "SWIR/MWIR InP-Based p-i-n Photodiodes with InGaAs/GaAsSb Type-II Quantum Wells," IEEE J. Quantum Electron. **30**, 399–402 (201AD).
17. B. Chen, W. Y. Jiang, J. Yuan, A. L. Holmes, and B. M. Onat, "Demonstration of a Room-Temperature InP-Based Photodetector Operating Beyond 3 μm," IEEE Photonics Technol. Lett. **23**, 218–220 (2011).
18. B. Chen and A. L. Holmes, "InP-based short-wave infrared and midwave infrared photodiodes using a novel type-II strain-compensated quantum well absorption region.," Opt. Lett. **38**, 2750–2753 (2013).
19. B. Chen, W. Y. Jiang, A. L. Holmes, and W. Y. J. A. L. Holmes, "Design of strain compensated InGaAs/GaAsSb type-II quantum well structures for mid-infrared photodiodes," Opt. Quantum Electron. **44**, 103–109 (2012).
20. B. Tossoun, R. Stephens, Y. Wang, S. Addamane, G. Balakrishnan, A. Holmes, and A. Beling, "High-Speed InP-Based p-i-n Photodiodes With InGaAs/GaAsSb Type-II Quantum Wells," IEEE Photonics Technol. Lett. **30**, 399–402 (2018).
21. Y. Chen, X. Zhao, J. Huang, Z. Deng, C. Cao, Q. Gong, and B. Chen, "Dynamic model and bandwidth characterization of InGaAs/GaAsSb type-II quantum wells PIN photodiodes," Opt. Express **26**, (2018).
22. B. Tossoun, J. Zang, S. J. Addamane, G. Balakrishnan, A. L. Holmes, and A. Beling, "InP-Based Waveguide-Integrated Photodiodes With InGaAs/GaAsSb Type-II Quantum Wells and 10-GHz Bandwidth at 2 μm Wavelength," J. Light. Technol. **36**, 4981–4987 (2018).
23. Y. Chen and B. Chen, "Design of InP-based high-speed photodiode for 2-μm wavelength application," IEEE J. Quantum Electron. **55**, (2019).
24. J. M. Wun, Y. W. Wang, and J. W. Shi, "Ultrafast Uni-Traveling Carrier Photodiodes with GaAs0.5Sb0.5/In0.53Ga0.47As Type-II Hybrid Absorbers for High-Power Operation at THz Frequencies," IEEE J. Sel. Top. Quantum Electron. **24**, 1–7 (2018).
25. C. C. Renaud, M. Natrella, C. Graham, J. Seddon, F. Van Dijk, and A. J. Seeds, "Antenna integrated THz uni-traveling carrier photodiodes," IEEE J. Sel. Top. Quantum Electron. **24**, (2018).
26. A. Beling, X. Xie, and J. C. Campbell, "High-power, high-linearity photodiodes," Optica **3**, 328–338 (2016).
27. T. Ishibashi, Y. Muramoto, T. Yoshimatsu, and H. Ito, "Unitraveling-Carrier Photodiodes for Terahertz Applications," IEEE J. Sel. Top. Quantum Electron. **20**, (2014).